# Demkov-Kunike model in cold molecule formation

A. Ishkhanyan

Institute for Physical Research, NAS of Armenia, Ashtarak 0203, Armenia

**ABSTRACT**

We study the dynamics of cold molecule formation via photo- or magneto-association of quantum degenerate atomic gases for the case when the field configuration is defined by the quasi-linear level crossing Demkov-Kunike model, which is characterized by a bell-shaped pulse and finite variation of the frequency detuning. We generalize the approach developed for the Landau-Zener model and propose a cubic polynomial equation for the asymptotic transition probability at $t \to +\infty$ applicable in the large detuning regime of the interaction process. The proposed approximation applies to all values of the Rabi frequency; hence, it presents a unified description of previously noticed weak, moderate, and strong interaction limits of the fast sweep regime. This cubic equation for the transition probability provides improvement of previous results by an order of the magnitude.

## 1. INTRODUCTION

Production of cold molecules by means of photo- or magneto-association of pre-existing ultracold atoms in Bose-Einstein condensates and degenerate Fermi gases has gained considerable interest during the last years (see, e.g., [1,2] and references therein). From the theoretical point of view, one of the tasks arising here is the effective control of non-adiabatic transitions [3,4] by time-dependent optical and magnetic fields in nonlinear quantum systems. To explore this point, several pulse configurations have been discussed. The simplest and widely used model is the Landau-Zener configuration [5,6] (a model assuming constant coupling and linear crossing of energy levels). Using this model, many qualitative and quantitative features of non-adiabatic transitions have been studied both in the context of linear and nonlinear dynamics (see, e.g., [3-18]). However, from the physical point of view, this model suffers several shortcomings: the field is never turned off, the model assumes infinite energy of the field at $t \to \pm\infty$, and the detuning varies strongly linearly in time. It is then understood that this model, though being a good approximation in the vicinity of the resonance crossing point, is not realizable physically. Hence, examination of models that do not suffer of mentioned disadvantages will result in more realistic representations on the behavior of the quantum system. One such a model is the first Demkov-Kunike model [19,20], which is an experimentally realizable version of a field configuration with finite-duration pulse and nearly-linear passage of the resonance. This is a rich model that displays several peculiarities different from those suggested by the Landau-Zener model. For instance, discussing the possible characteristic cases of cold atom photoassociation under the strong nonlinearity conditions, it has been shown that in general two qualitatively distinct behavioral regimes occur [21,22]. In the case of large frequency detuning and high field intensities, the molecule formation process takes place with only weakly expressed oscillations between the atomic and molecular populations, while in the second case, when the detuning is small, the system displays large-amplitude Rabi-type oscillations between the two populations [21,22]. It is worth stressing that in the case of the Landau-Zener model only one of these two regimes is revealed; in that case, the system always follows the weekly oscillatory scenario.

A convenient framework to consider the non-adiabatic transitions in nonlinear quantum systems is the semi-classical two state problem. Different nonlinear extensions of familiar linear two-state problem have been formulated and discussed during the last decades in connection with, e.g., second harmonic generation in nonlinear optics, physics of interacting Bose-condensates, cold atom association in Bose-condensates and degenerate Fermi-gases, etc. Here, we consider a basic extension of the two-state problem to the *quadratic-nonlinear* case [23-25] that is generic for the field theories involving a Fermi 2:1 resonance:

$$i\frac{da_1}{dt} = U(t)e^{-i\delta(t)}\overline{a}_1 a_2, \qquad (1)$$

$$i\frac{da_2}{dt} = \frac{U(t)}{2}e^{+i\delta(t)}a_1 a_1. \qquad (2)$$

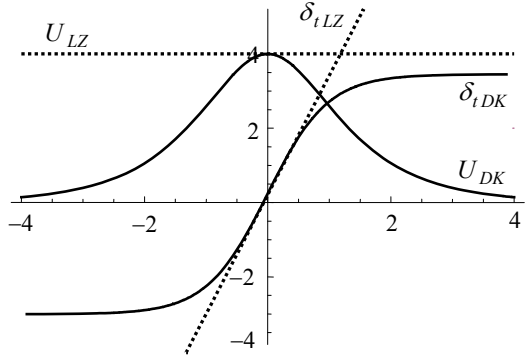

Fig. 1. The Demkov-Kunike configuration (dotted lines - the Landau-Zener model).

In the context of the photoassociation of an atomic Bose-Einstein condensate, this is the model system of coupled mean-field Gross-Pitaevskii-type nonlinear equations describing atomic and molecular condensates as classical fields. Here, $a_1$ and $a_2$ are the atomic and molecular states' probability amplitudes, respectively, $U(t)$ is the Rabi frequency of the associating laser field, $\delta(t)$ is the detuning modulation function, which is defined as the integral from the detuning $\delta_t$ of the laser frequency from the frequency of the transition in which two atoms are converted into a molecule. The Demkov-Kunike model we discuss is [19,20]

$$U(t) = U_0 \operatorname{sech}(t), \quad \delta_t(t) = 2\delta_0 \tanh(t). \tag{3}$$

This field configuration defines a bell-shaped laser pulse with finite variation of the frequency detuning, which quasi-linearly crosses the resonance, i.e., $\delta_t = 0$, at the time point $t = 0$ (Fig. 1). This model has been previously addressed in several publications. The weak interaction limit, corresponding to small values of the Rabi frequency, has been discussed in [26,27]. Applying Picard's successive approximations to an exact nonlinear Volterra integral equation [26] or a specific variational approach involving the solution of the associated linear problem [27], analytical approximations for the probability of transition to the molecular state at $t \to +\infty$ have been obtained. The strong coupling limit, corresponding to large values of Rabi frequency, has been addressed in [21,22,28,29]. In [21,22], it has been shown that this interaction limit is effectively subdivided into two different interaction regimes corresponding to slow and fast sweep through the resonance, i.e., to small and large detuning, respectively. When the passage through the resonance is slow ($\delta_0 < 1$), the system exhibits large-amplitude Rabi-type oscillations between atomic and molecular states' populations (Fig. 2, a)). In the opposite regime, i.e., in the case of fast enough crossing of the resonance ($\delta_0 > 1$, only weak, damped oscillations between the involved two states are observed (Fig. 2, b)).

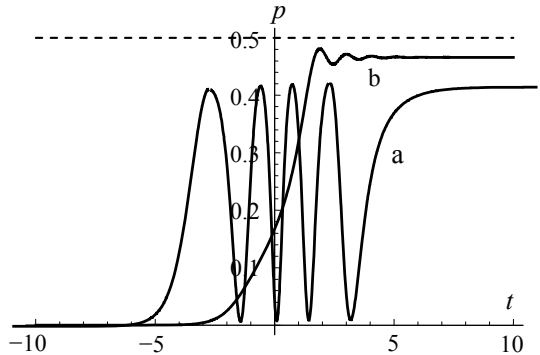

Fig. 2. Transition probability versus time. a) Small detuning regime of the strong coupling limit ($U_0 = 8$, $\delta_0 = 0.1$). b) Large detuning regime of the strong coupling limit ($U_0 = 8$, $\delta_0 = 8$).

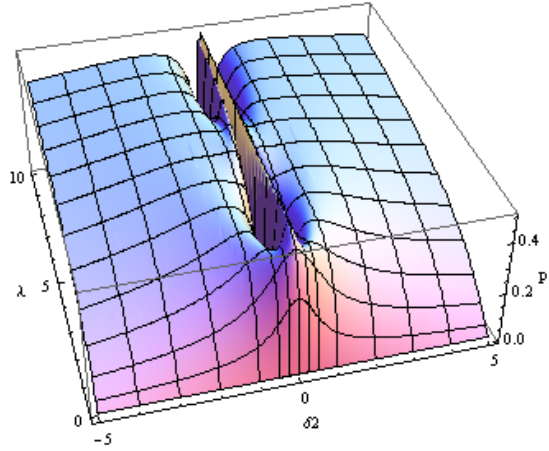

Fig. 3. The dependence of the transition probability at $t \to +\infty$ on the input parameters $\delta_0$ and $\lambda = U_0^2$.

These two strongly nonlinear interaction regimes occurring at small and large detuning cases display qualitatively distinct dependence of the *asymptotic transition probability* at $t \to +\infty$ on the input parameters of the problem $U_0$ and $\delta_0$. This is illustrated in Fig. 3. It is seen that at small $\delta_0 < 1$ the dependence $p(+\infty, U_0)$ is oscillatory, while at large detuning $\delta_0 > 1$ the dependence is almost monotonic. The strong interaction limit of the large detuning regime has been studied in [28,29]. In [28], an approximate solution has been constructed which is defined as a solution of a first-order nonlinear differential equation and contains a fitting parameter, which is determined through a variational procedure. However, this approximation misses several essential features of the association process such as the coherent oscillations between atomic and molecular populations which arise after the system passes through the resonance. In [29], the applied approach has been further developed and the next approximation to the problem has been constructed using the previous solution as a zero-order approximation. The resultant two-term approximation, that now describes the mentioned oscillations, contains two more fitting parameters that are determined combining analytical and numerical methods. Finally, unified analytic approximations for the involved variational parameters have been proposed. With these choices of parameters, the constructed approximation becomes equally applicable to weak, moderate and strong coupling cases of the large detuning regime of the process.

In the present paper we discuss the problem using a different approach based on the analysis of the asymptotic behavior of the nonlinear system at $t \to +\infty$ and comparing it with that for corresponding linear solution. The approach allows one to derive a cubic equation for the efficiency of conversion of atoms into molecules. The equation has a simple structure and involves the transition probability of the associated linear problem as a parameter. A root of the proposed cubic equation improves the previous result by an order of the magnitude.

## 2. TREATMENT AND RESULTS

It has been shown that the time dynamics of the molecular state probability $p = |a_2|^2$ is described by the following exact nonlinear ordinary differential equation of the third order [21-22]:

$$\left(\frac{d}{dt} - \frac{1}{\delta_t}\frac{d\delta_t}{dt}\right)\left[\frac{1}{U}\frac{d}{dt}\left(\frac{1}{U}\frac{dp}{dt}\right) - \frac{1}{2}(1 - 8p + 12p^2)\right] + \delta_t^2 \frac{dp}{dt} = 0. \quad (4)$$

Denoting

$$F = \frac{1}{U}\frac{d}{dt}\left(\frac{1}{U}\frac{dp}{dt}\right) - \frac{1}{2}(1 - 8p + 12p^2), \quad (5)$$

this equation is rewritten as

$$\frac{d}{dt}\left(\frac{F}{\delta_t}\right) + \delta_t \frac{dp}{dt} = 0 \quad (6)$$

or
$$\frac{F}{\delta_t} + \int \delta_t \frac{dp}{dt} dt = 0 . \tag{7}$$

On the other hand, multiplying by $F/\delta_t$, Eq. (6) can be rewritten as

$$\frac{F}{\delta_t} \frac{d}{dt}\left(\frac{F}{\delta_t}\right) + F\frac{dp}{dt} = 0 , \tag{8}$$

so that

$$\left(\frac{F}{\delta_t}\right)^2 + 2\int F\frac{dp}{dt} dt = 0 . \tag{9}$$

Combining now Eqs. (7) and (9), we obtain

$$2\int F dp = \left(\int \delta_t \frac{dp}{dt} dt\right)^2 . \tag{10}$$

The integration in the left-hand side is readily performed yielding

$$\left(\frac{1}{U}\frac{dp}{dt}\right)^2 - p(1-2p)^2 = \left(\int \delta_t \frac{dp}{dt} dt\right)^2 . \tag{11}$$

This integro-differential equation serves as the basis for further developments.

As the next step, we examine an auxiliary linear problem (compare with Eqs. (1),(2)):

$$i\frac{da_{1L}}{dt} = U_1(t)e^{-i\delta(t)} a_{2L} , \tag{12}$$

$$i\frac{da_{2L}}{dt} = U_1(t)e^{+i\delta(t)} a_{1L} , \tag{13}$$

with variable normalization $|a_{1L}|^2 + |a_{2L}|^2 = I$ and changed Rabi frequency $U_1(t)$. The second state's probability $p_L = |a_{2L}|^2$ for this system obeys the equation (compare with Eq. (4))

$$\left(\frac{d}{dt} - \frac{1}{\delta_t}\frac{d\delta_t}{dt}\right)\left[\frac{1}{U_1}\frac{d}{dt}\left(\frac{1}{U_1}\frac{dp_L}{dt}\right) - \frac{1}{2}(4I - 8p_L)\right] + \delta_t^2 \frac{dp_L}{dt} = 0 , \tag{14}$$

From this equation we derive an equation that presents the analogue of equation (11) for the linear problem (12)-(13):

$$\left(\frac{1}{U_1}\frac{dp_L}{dt}\right)^2 - p_L(4I - 4p_L) = \left(\int \delta_t \frac{dp_L}{dt} dt\right)^2 . \tag{15}$$

Now, in order to meet the normalization of the nonlinear problem, $|a_1|^2 + 2|a_2|^2 = 1$, we choose $I = 1/2$, so that $|a_{2L}|^2 \in [0, 1/2]$. Further, we put $U_1(t) = C_0 U(t)$ and consider a specific linear problem that satisfies the initial conditions of the nonlinear problem and, additionally produces the same transition probability at $t \to +\infty$ as the nonlinear problem: $p_L(+\infty) = p(+\infty)$. The latter condition formulates a boundary value problem, the eigenvalue of which defines the appropriate value of $C_0$. It is now understood that this auxiliary linear solution has the same asymptotes as the nonlinear solution. Consider now the transition probability at $t \to +\infty$. Since $p_L(+\infty) = p(+\infty) \equiv p_{\text{inf}}$, dividing Eq. (15) by 2 and subtracting Eq. (11) we get

$$2p_{\text{inf}}^2(1-2p_{\text{inf}}) = \left\{\left(\left(\int_{-\infty}^t \delta_t \frac{dp}{dt} dt\right)^2 - \frac{1}{2}\left(\int_{-\infty}^t \delta_t \frac{dp_L}{dt} dt\right)^2\right) - \left(\left(\frac{1}{U}\frac{dp}{dt}\right)^2 - \frac{1}{2}\left(\frac{1}{U_1}\frac{dp_L}{dt}\right)^2\right)\right\}\Bigg|_{+\infty} . \tag{16}$$

Since the nonlinear solution $p(t)$ and the auxiliary linear solution $p_L(t)$ have the same asymptotes at $t \to +\infty$, it is expected that the right-hand side of this equation is small. On the other hand, it is proportional to $1/U_0^2$. Hence, it will diverge at the limit of vanishing field intensities, $U_0^2 \to 0$, unless it is proportional to $(p_{\text{inf}} - p_L/2)$. This is because it is established that in this limit $p_{\text{inf}} \sim p_L/2$ [12]. We stress that for the specific linear solution under consideration the normalization $I = 1/2$ is applied. So that, $p_L / 2 = p_{DK} / 4$, where $p_{DK}$ is the familiar solution to the linear Demkov-Kunike problem with normalization $I = 1$:

$$p_{DK} = 1 - \cos^2\left(\pi\sqrt{U_0^2 - \delta_0^2}\right)\operatorname{sech}^2(\pi\delta_0). \tag{17}$$

Thus, we are lead to the following cubic equation for the transition probability at the end of the process:

$$2p_{\inf}^2(1 - 2p_{\inf}) = \frac{a}{U_0^2}\left(p_{\inf} - \frac{p_{DK}}{4}\right), \tag{18}$$

where $a$ is a constant of the order of $O(1)$. This is an advanced approximation. Numerical simulations show, that a root of this equation describes the asymptotic transition probability at $t \to +\infty$ with accuracy of the order of $10^{-4}$. This result improves the previous analytic approximation of [29] by the order of the magnitude. The graphs produced by this solution are compared with the numerical result in Fig. 4, where successive sections of Fig. 3 for different $\delta_0 > 1$ are presented. An important observation coming out from these graphs is that the transition probability for a nonzero $\delta_0$ never riches $1/2$, but is restricted by a limiting value $p = p_{sat} < 1/2$. Thus, in contrast to the linear case, in the nonlinear case complete conversion of atoms into molecules is impossible for a Demkov-Kunike resonance crossing process.

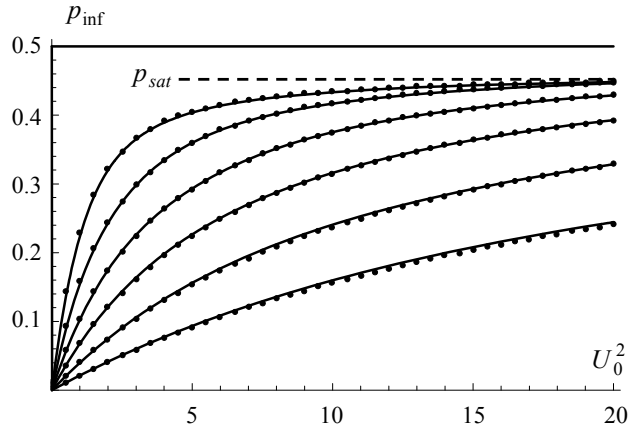

Fig. 4. Asymptotic transition probability at $t \to +\infty$ versus $U_0^2$ at $\delta_0 = 2, 3.5, 6, 10, 18, 35$ (from left to right). Filled circles – numerical result, solid line- the root of Eq. (18). The nonlinear saturation of the transition is shown by the dashed line $p = p_{sat}$.

## 3. CONCLUSIONS

Thus, we have discussed the photo- or magneto-association of ultracold atoms into molecules by means of a Demkov-Kunike quasi-linear resonance crossing process. This model, which is characterized by a finite-duration bell-shaped pulse and finite variation of the frequency detuning (if the photoassociation terminology is used), is an experimentally realizable version of a field configuration that does not suffer the known disadvantages of the linear-in-time term-crossing model by Landau and Zener. Discussing the transition probability at $t \to +\infty$, we have proposed a cubic polynomial equation that applies to all values of the Rabi frequency as far as the large detuning regime of the interaction process is considered. This cubic equation thus suggests a unified description of previously noticed weak, moderate, and strong coupling limits of the fast sweep regime. Notably, the equation provides improvement of previous results by an order of the magnitude. The proposed approximation shows that, in contrast to the linear case, in the nonlinear case complete conversion of atoms into molecules is impossible for a Demkov-Kunike resonance crossing process.

## ACKNOWLEDGMENTS

This research has been conducted within the scope of the International Associated Laboratory IRMAS. The work was supported by the Armenian National Science and Education Fund (ANSEF Grant No. 2010-PS-2186).